\documentclass[3p,times,procedia]{elsarticle}

\usepackage{ecrc}
\volume{00}

\firstpage{1}

\journalname{Physics Procedia}

\runauth{M. Soares-Santos et al.}

\jid{phpro}

\jnltitlelogo{Physics Procedia}

\CopyrightLine{2011}{Published by Elsevier Ltd.}

\usepackage{amssymb}
\usepackage[figuresright]{rotating}

\begin{document}

\begin{frontmatter}

%% Title, authors and addresses

%% use the tnoteref command within \title for footnotes;
%% use the tnotetext command for the associated footnote;
%% use the fnref command within \author or \address for footnotes;
%% use the fntext command for the associated footnote;
%% use the corref command within \author for corresponding author footnotes;
%% use the cortext command for the associated footnote;
%% use the ead command for the email address,
%% and the form \ead[url] for the home page:
%%
%% \title{Title\tnoteref{label1}}
%% \tnotetext[label1]{}
%% \author{Name\corref{cor1}\fnref{label2}}
%% \ead{email address}
%% \ead[url]{home page}
%% \fntext[label2]{}
%% \cortext[cor1]{}
%% \address{Address\fnref{label3}}
%% \fntext[label3]{}

\dochead{\small{TIPP2011 -- 2nd International Conference on Technology and Instrumentation in Particle Physics}}
%% Use \dochead if there is an article header, e.g. \dochead{Short communication}
%% \dochead can also be used to include a conference title, if directed by the editors
%% e.g. \dochead{17th International Conference on Dynamical Processes in Excited States of Solids}

\title{DECam integration tests on telescope simulator}

%% use optional labels to link authors explicitly to addresses:
%% \author[label1,label2]{<author name>}
%% \address[label1]{<address>}
%% \address[label2]{<address>}

\author[label1]{M. Soares-Santos\corref{cor1}}
\cortext[cor1]{Corresponding author: {\tt marcelle@fnal.gov}}
\author[label1]{J. Annis}
\author[label6]{M. Bonati}
\author[label1]{E. Buckley-Geer} 
\author[label1]{H. Cease}
\author[label5]{D. DePoy}
\author[label1]{G. Derylo}
\author[label1]{H. T. Diehl}
\author[label2]{A. Elliott}
\author[label1]{J. Estrada} 
\author[label1]{D. Finley}
\author[label1]{B. Flaugher}
\author[label1]{J. Frieman}
\author[label1]{J. Hao}
\author[label2]{K. Honscheid}
\author[label3]{I. Karliner}
\author[label1]{K. Krempetz}
\author[label4]{K. Kuehn}
\author[label4]{S. Kuhlmann}
\author[label1]{K. Kuk}
\author[label1]{H. Lin}
\author[label1]{W. Merrit}
\author[label1]{E. Neilsen}
\author[label1]{L. Scott}
\author[label7]{S. Serrano}
\author[label1]{T. Shaw}
\author[label1]{K. Schultz}
\author[label1]{W. Stuermer}
\author[label8]{A. Sypniewski}
\author[label3]{J. Thaler}
\author[label6]{A. Walker}
\author[label3]{J. Walton}
\author[label1]{W. Wester}
\author[label1]{B. Yanny}

\address{FOR THE DARK ENERGY SURVEY COLLABORATION}

\address[label1]{Fermi National Accelerator Laboratory}
\address[label6]{National Optical Astronomy Observatory}
\address[label5]{Texas A \& M University}
\address[label2]{Ohio State University}
\address[label3]{University of Illinois at Urbana-Champaign}
\address[label4]{Argonne National Laboratory}
\address[label7]{Institut d'Estudis Espacials de Catalunya}
\address[label8]{University of Michigan}

\begin{abstract}
%% Text of abstract
The Dark Energy Survey (DES) is a next generation optical survey aimed at measuring the expansion history of the 
universe using four probes: weak gravitational lensing, galaxy cluster counts, baryon acoustic oscillations, and Type 
Ia supernovae. To perform the survey, the DES Collaboration is building the Dark Energy Camera (DECam), a 3 
square degree, 570 Megapixel CCD camera which will be mounted at the Blanco 4-meter telescope at the Cerro 
Tololo Inter-American Observatory. DES will survey 5000 square degrees of the southern galactic cap in 5 filters 
(g, r, i, z, Y). DECam will be comprised of 74 250 micron thick fully depleted CCDs: 62 2k x 4k CCDs for imaging 
and 12 2k x 2k CCDs for guiding and focus. Construction of DECam is nearing completion. In order to verify that 
the camera meets technical specifications for DES and to reduce the time required to commission the instrument, we 
have constructed a full sized telescope simulator and performed full system testing and integration prior to shipping. 
To complete this comprehensive test phase we have simulated a DES observing run in which we have collected 4 
nights worth of data. We report on the results of these unique tests performed for the DECam and its impact on the 
experimentÕs progress.
\end{abstract}

\begin{keyword}
%% keywords here, in the form: keyword \sep keyword
Dark Energy \sep CCD \sep camera \sep survey
%% PACS codes here, in the form: \PACS code \sep code

%% MSC codes here, in the form: \MSC code \sep code
%% or \MSC[2008] code \sep code (2000 is the default)

\end{keyword}

\end{frontmatter}

%%
%% Start line numbering here if you want
%%
% \linenumbers

%% main text
\section{Introduction}\label{Introduction}
The Dark Energy Survey (DES, {\tt darkenergysurvey.org}) 
\cite{DES:2005,Annis:2005,Annis:2005a} is a ground-based photometric survey conceived to shed light on the problem of the accelerated expansion of the universe by measuring the dark energy equation of state parameter with four complementary techniques: galaxy cluster counts, weak lensing, galaxy angular power spectrum and Type Ia supernovae. The combination of these techniques will result in a factor of $\sim$3-5 improvement over current experiments in the figure of merit defined by the Dark Energy Task Force \cite{Albrecht:2006}. This will be achieved by measuring, in the redshift range $0 \lesssim z \lesssim 1.5$, redshifts and shapes of 300 million galaxies, mass and redshifts of tens of thousands of galaxy clusters, luminosity and redshifts of about 4000 Type Ia supernovae. DES will survey a 5000 square-degrees area of the sky in the optical grizY bands and its total volume, estimated to be of 24 $h^{-3}$Gpc$^{3}$, %\cite{Banerji:2008}, 
is 7 times larger than the largest existing CCD survey of the Universe by volume to date, the Sloan Digital Sky Survey Luminous Red Galaxies photometric sample \cite{Thomas:2011}. 
%Details on the DES expected constraints and systematic uncertainties can be found in our white papers \cite{DES:2005,Annis:2005,Annis:2005a}. 
DES  will also provide  the astronomical community with a  wide  field, 5 band digital survey of  the  southern sky with excellent image quality, uniform  photometry and unprecedented depth for its sky coverage.
%The DES survey area overlaps with that of several other important surveys for example the southern equatorial stripe of the SDSS and the South Pole Telescope Sunnyaev-Zeldovich cluster survey and the infra-red surveys being conducted on the Visible and Infra-Red Survey Telescope for Astronomy (VISTA) at ESOÕs Cerro Paranal Observatory in Chile.

To accomplish our science goals the DES collaboration has designed \cite{Flaugher:2006,Cease:2008} and built 
\cite{Derylo:2010,Flaugher:2010} the Dark Energy Camera (DECam) \cite{DePoy:2008,Honscheid:2008}, a new imaging instrument comprised with 74 250 micron thick CCDs \cite{Holland:2003,Estrada:2010}. DECam will be installed on the Cerro Tololo Inter-American Observatory (CTIO) 4-meter Blanco telescope \cite{Abbott:2006}. DES will use 525 nights of DECam observations carried in the 2012-2017 austral spring and summer months. 

Construction of DECam is nearing completion. In order to verify that the camera meets technical specifications for DES and to reduce the time required to commission the instrument, we have constructed a full sized telescope simulator \cite{Diehl:2010} and performed full system testing and integration prior to shipping. In this paper we report on the results of these unique tests performed for the DECam and its impact on the experiment's progress. Updated overviews of the DECam project \cite{Diehl:2011}, the CCD properties \cite{Kubik:2011} and readout electronics \cite{Shaw:2011} can be found in this conference proceedings volume. 

This paper is organized as follows: Section \S\ref{Instrument} describes the DECam hardware components 
available for testing on the telescope simulator (\S\ref{DECam}) and the data acquisition system (\S\ref{SISPI}).
Section \S\ref{Tests} details the telescope simulator setup, goals and methods.
Section \S\ref{Data} presents the data taken during this test phase. 
Our results are discussed in section \S\ref{Results} and followed by a brief Conclusions section, \S\ref{Conclusion}.

\section{Instrument}\label{Instrument}

\subsection{DECam}\label{DECam}
\begin{figure}[h!]
\centering
\includegraphics[width=0.49\linewidth]{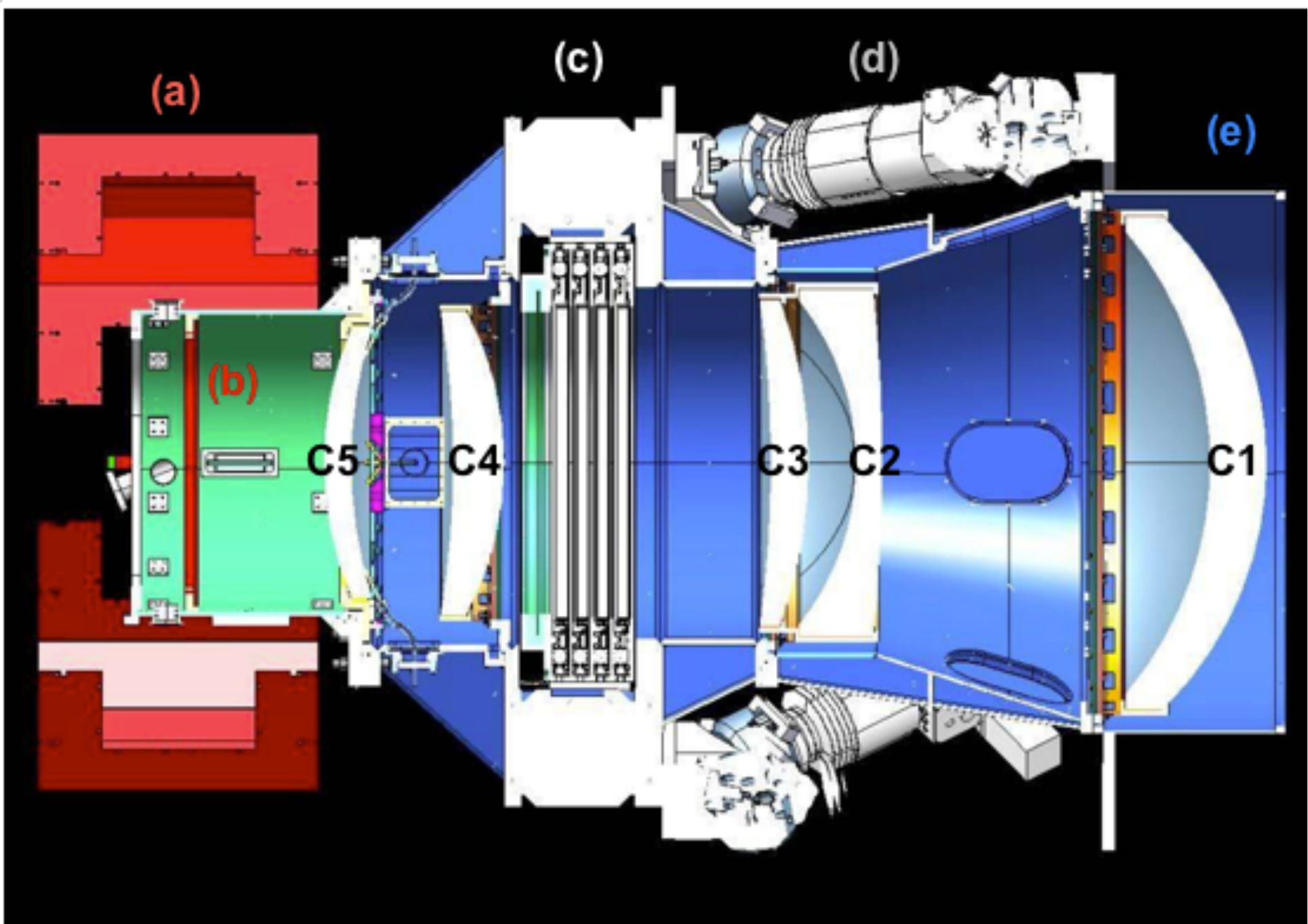} \hspace{0.05\linewidth}
\includegraphics[width=0.44\linewidth]{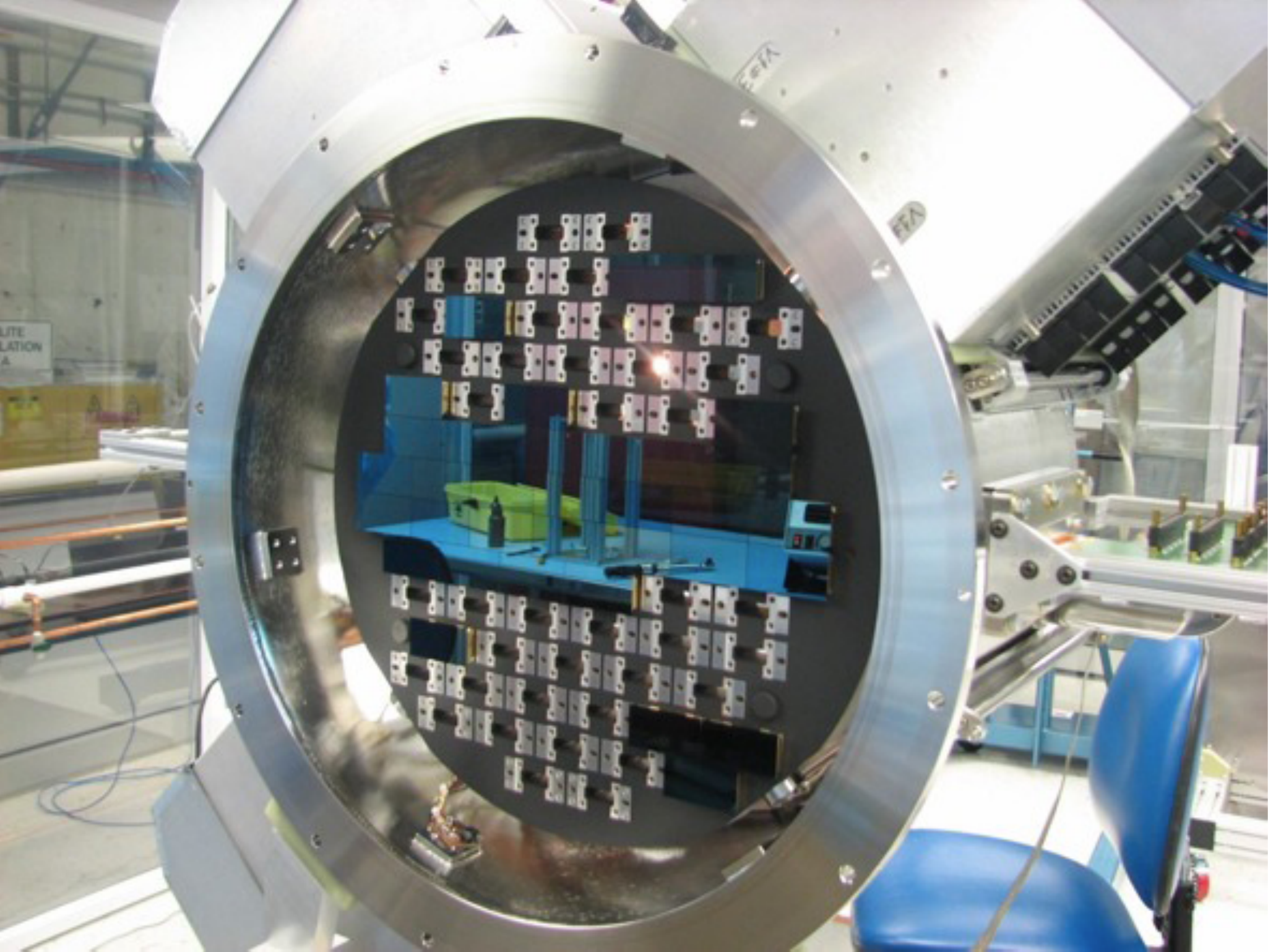}
\caption{Left: DECam schematic view showing the components tested on the telescope simulator: (a) readout electronics crates, (b) imager focal plane, (c) filter changer and shutter system and (d) hexapod. 
The optical corrector lenses, C1 to C5, and the barrel (e) are also shown.
Right: Focal plane populated with 24 engineering grade CCDs for testing at the telescope simulator.}
\label{overview}
\end{figure}

DECam is a new imaging instrument for the Blanco 4-meter telescope at CTIO. Its major components are: a 570 Megapixel cryocooled \cite{Cease:2010} optical CCD imager with 74 250 microns thick detectors \cite{Estrada:2010}  
that were developed at LBNL \cite{Holland:2003} and packaged at Fermilab 
\cite{Derylo:2006,Estrada:2006,Diehl:2008,Kubik:2010,Kubik:2011}, a low noise readout
system \cite{Castilla:2010,Shaw:2010,Shaw:2011} housed in actively cooled crates, a two-blade shutter, a filter changer with 8 slots (5 of which used for DES filters), a hexapod for real-time focus and alignment with precision of 1-micron and an optical corrector with 5 lenses \cite{Kent:2006}. Of the 74 CCDs, 62 are 4k x 2k chips used for science imaging, 12 are 2k x 2k of which 4 are used for guiding and 8 for focus. 
Figure~\ref{overview} shows these main components, highlighting those that were 
available for testing at the telescope simulator: (a) the readout crates, (b) the imager focal plane, 
(c) shutter and filter changer, (d) hexapod. Also shown in the picture are the corrector 
lenses and the barrel. The focal plane was loaded with 27 engineering grade CCDs (24 imaging, 3 guiding detectors) as shown in Fig.~\ref{overview}. 
Tests with the final set of 74 science grade CCDs currently installed on the focal plane were performed after we were satisfied with the results on the telescope simulator. We did not include any of the optics components in our tests. Both lenses and filters were tested independently and are being shipped directly to CTIO. We used dummy weights in the place of the optical components and a flat optical window instead of C5 for our tests. 

\subsection{Data acquisition system}\label{SISPI}
Survey Image System Process Integration (SISPI) is the DES data acquisition system \cite{Honscheid:2008a}. 
Image data from the focal plane CCDs, read by the Monsoon front end electronics, flow to the 
Image Acquisition and Image Builder systems before being recorded on a storage device and handed over to the DES 
Data Management system \cite{Mohr:2008}. The data flow is coordinated by the Observation Control system (OCS) which determines 
the exposure sequence and is assisted by the Instrument Control (ICS) and Image Stabilization systems. Pointing 
information, correction signals derived from the guide CCDs and other telemetry information is exchanged with the 
Blanco Telescope Control system (TCS). The ICS implements hardware control loops for the shutter, filter changer and 
cooling system. 
With 62 CCDs or 520 Mega pixels and 16 bits per pixel, one DECam exposure generates approximately 1 GByte of data. At a rate of 250 kpix/s it takes about 17 seconds to transfer the data from the focal plane to the computers of the Image Acquisition system. During this time the telescope slews to the next position and SISPI must be ready for the 
next exposure. The images transfer via Gigabit Ethernet links to the Image Building system where it is packaged in multi-extension FITS files and stored on disk. The SISPI applications are built upon a common software infrastructure layer which provides the message passing system, the facility database and support for configuration and initialization.
In SISPI we distinguish between Command messages and Telemetry data. 
Commands are used to request information from a remote application or to activate a remote 
action. The Command or Message Passing system is implemented using a Client-Server design pattern based the 
CTIO developed Soar Messaging Library (SML). Telemetry in this context refers to any information an application 
provides and that can be of use to another application. In order to provide this functionality we implemented a 
Shared Variable system (SVE) using the Publish-Subscribe design pattern. The read-out and control system are operational and were  tested on the telescope simulator. 

\section{Telescope Simulator: concept, goals and methods}\label{Tests}

\begin{figure}[h!]
\centering
\includegraphics[width=0.49\linewidth]{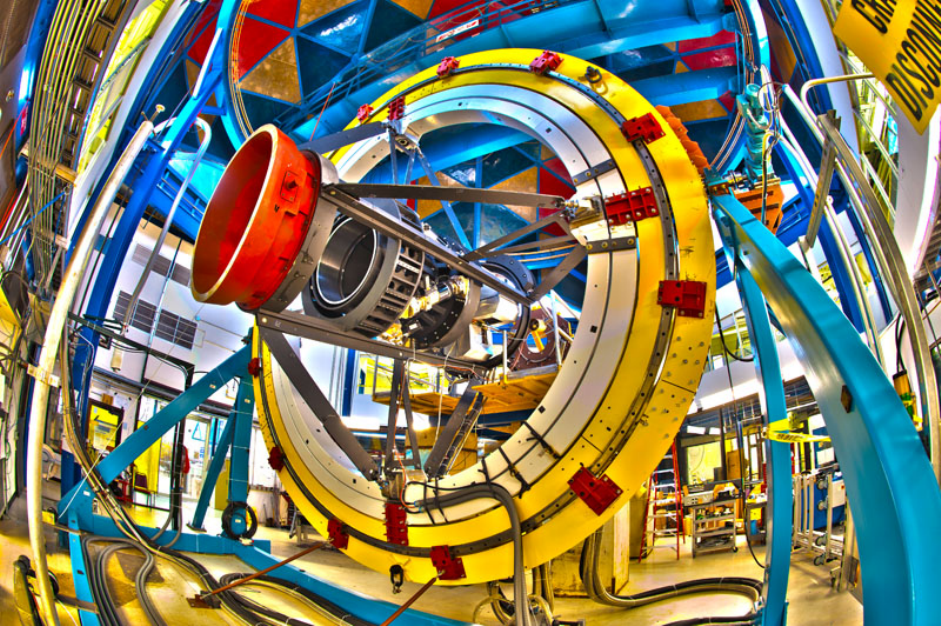}\hspace{0.05\linewidth}
\includegraphics[width=0.44\linewidth]{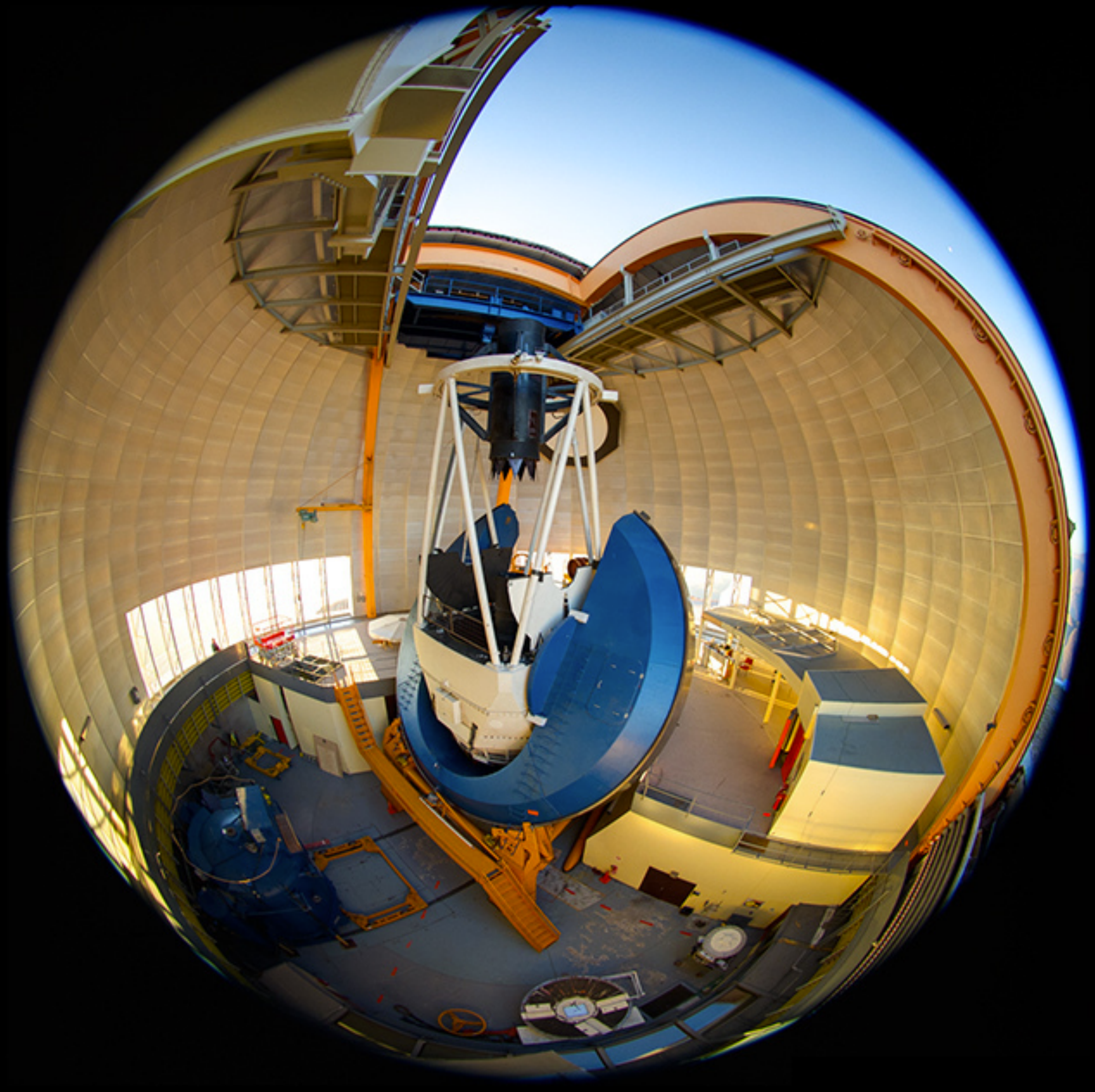}
\caption{Left: DECam on the telescope simulator (Photo credit: Reidar Rahn, Fermilab). 
Right: Blanco telescope (Photo credit: Jose Francisco Salgado,Adler Planetarium). 
The two white rings on the telescope simulator replicate the Blanco top end rings.}
\label{simulator}
\end{figure}

The Telescope Simulator \cite{Diehl:2010}, shown in Fig.~\ref{simulator}  is a full-size replica of the Blanco telescope top end rings and the ÒfinsÓ that connect the camera to those rings. 
The camera, fins, and rings are supported by a frame that provides pitch and roll capability to orient 
that equipment as if it were at the top of a telescope. 
Though the machine has the capability to move, it does not mimic the 
slewing or guiding capability of the telescope, which means that our tests were performed with DECam 
statically positioned at different orientations.  
The telescope simulator  is 14 ft tall and 25 ft wide. The four rings weigh 
32,000 pounds. The outer one has a 24 ft diameter. 
Two motors from SEW Eurodrive can orient the camera to any angle 
within the pitch and roll degrees of freedom.   The pitch motor is 1/3 HP, 1800 RPM, directly-coupled, torque-limited 
and geared down  35,009:1 for a maximum speed of 20 minutes per revolution. The roll motor is 1/2 HP, 1800 RPM, 
geared down 709:1 for a maximum speed of 11.5 minutes per revolution.  The coupling is by means of a 62 ft chain 
attached to the inner race (3rd ring out).   
When the assembly is not moving the motors automatically engage brakes. The motors are controlled from a panel 
located on the exterior of the base.  These controls are simple power on/off, with forward/backward and speed control 
for each ring. Four limit switches prevent the rings from being oriented in any undesired direction.  
A Star and Flat Field Projector \cite{Hao:2010} is mounted at the end of the cage and is used to illuminate the focal plane. 
In the flat field configuration, the Projector consists of a laser and an integrating sphere. The light level can be adjusted
by changing the intensity of the beam. By replacing the sphere with a diffractive grid, we project point sources. 

%when was the TS built? how long did the test phase last?

%\subsection{Goals of testing}
The primary goals of the telescope simulator testing phase were to 
(1) verify that DECam meets all the requirements for acceptance by CTIO, 
(2) test and exercise the procedures for installing the new prime focus cage,
(3) establish and exercise the procedures to safely handle the instrument during installation at the mountain top and 
(4) minimize the time required for integration and commissioning. 
Specifically for SISPI, the main goal was to
(4) integrate the various components, verifying that the data acquisition system is fully operational.
In addition, with the telescope simulator setup at our disposal we had the opportunity to 
(5) obtain feedback about the system's readiness for commissioning from experienced observers within the collaboration, identifying improvements to be made prior to commissioning,
(6) develop and exercise some of the operational procedures to be used at the mountain top and train personnel to operate the system.

%
%\subsection{Methods}\label{Methods}
The methods employed in order to achieve these goals are listed below:
\begin{enumerate}
\item To develop and practice the procedures that will be used to install instrument on the telescope, the 
cage and fins, including the 
hexapod and barrel were inserted into the new Prime Focus Cage and 
the installed onto the ring system using a circular ceiling crane that is similar to that attached to the 
telescope dome. Then the DECam imager was installed onto the camera barrel 
using the imager installation/removal hardware that is being provided to CTIO. 
\item We verified that all systems at hand were operating according to their technical specifications and requirements
by operating the camera oriented in 6 positions: with the imager pointed down, as if the telescope was at zenith, and with the imager aimed near the horizon in 5 roll orientations covering 131 degrees. We took series of darks, flat fields
at different illumination levels and point source images. A summary of  our findings in comparison to the requirements 
is provided in the Results section \S\ref{Requirements}. 
\item Methods developed and exercised for installation and removal of components such as the shutter and filter changer will also help minimize the telescope down-time for commissioning at CTIO.
%\item Methods developed and 
%issues arisen during integration of the different components were solved and documented. This will help 
%minimize the telescope down-time for commissioning at CTIO.
\item SISPI components were integrated and tested as new hardware components were added to the system.
But the most comprehensive test was done by performing a simulated (mock) observing run, in which  we operated DECam as if we were taking DES data at the Blanco telescope. 
Results of the mock observing run are discussed in section \S\ref{MObs}.
Extensive feedback was collected from the observers during the mock observing run and summarized in a 
report. This helped to plan the preparatory work for commissioning. 
\item A draft version of an observer's guide was produced for the mock observing run and serves as a 
baseline for the document that will be provided to observers during DES operations.
\end{enumerate}

\section{Data}\label{Data}
\begin{figure}[h!]
\centering
\includegraphics[width=0.95\linewidth]{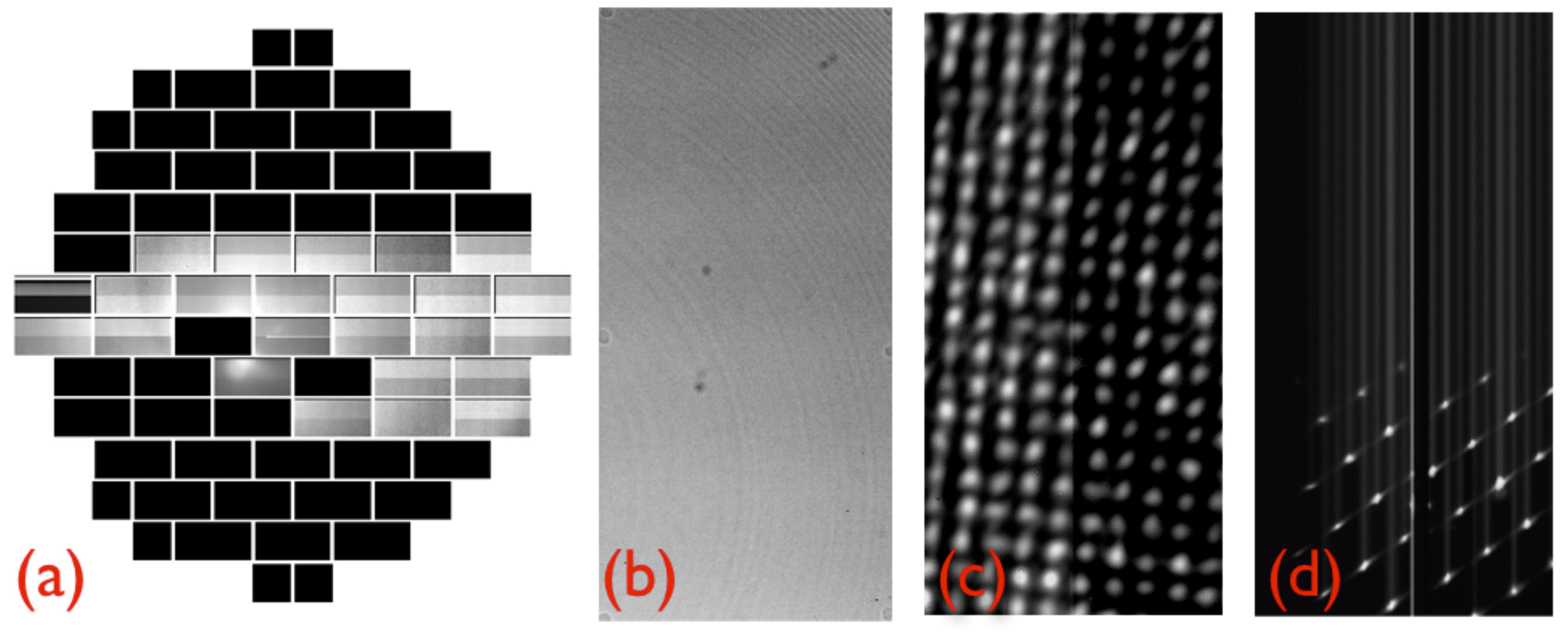}
\caption{Sample DECam data taken at the telescope simulator, showing the focal plane mosaic (a) 
with 24 CCDs installed %(see photograph on the left side of Fig.~\ref{overview} for comparison) 
and a 
series of single CCD images: (b) flat field, (c) projected point sources, (d) point sources with trails of light 
obtained by reading out the CCDs with the shutter open.}
\label{SampleImages}
\end{figure}

The imager was operated on the telescope simulator for about 5 months, from Oct 2010 to Feb 2011 and during this time, we took more than 10 thousand images of various types. The week of Feb 14-19 2011 was devoted to mock observing and 280 images were produced in that week.   
Examples of the images taken are shown in Fig.~\ref{SampleImages}. 
In addition to the image types expected in 
regular DES operations (biases, darks, flat fields and science fields) we also simulated ``focus'' images by performing  
traditional focus sweep sequences and ``trail'' images by reading out with the shutter open. The focus images were
used to test the hexapod movements, but DES will not rely on focus 
sequences to maintain focus. DECam has dedicated focus chips at the edges of the focal plane and will analyze their images to adjust the focus and alignment from one image to the next.
The trail images were used to test the vibrations requirements. 

\section{Results}\label{Results}

\subsection{Verification of technical requirements}\label{Requirements}
Data obtained during the testing of the imager on the telescope simulator were used to 
verify the DECam technical requirements \cite{Abbott:2009}. The tests were detailed in a report 
\cite{Estrada:2011} and a summary of the results is presented in Table \ref{RequirementsTable}. 
\begin{table}[h]
\centering
\begin{tabular}{l l l l}
\hline \\
ID &  DESCRIPTION & REQUIREMENT & MEASUREMENT\\\\
\hline \\
TG6 & Internal charged particle rate & $< 5$ events/cm$^{2}$/min & 2.7 events/cm$^{2}$/min\\
TG8 & Function in all orientations & Yes & Yes$^{(a)}$ \\
TG10 & Function in all expected environment & Yes & Yes$^{(b)}$\\
TO10 & Flat Focal Plane peak to peak variations & $<30\mu$m & $<15\mu$m $^{(c)}$\\
TD1 & Nonlinearity & $<1\%$ & $1\%$\\
TD2 & Full well capacity & $>130$k $e^{-}$ & $>130$k $e^{-}$ \\
TD4 & Readout rate & 17 seconds & 17 seconds\\
TD9 & CCD temperature stability with time& $<\pm0.25$K over 12 h & $<\pm0.15$K over 3 days \\
TD11 & CCD temperature stability across FP & $<10$C & $<8$C  \\
TD12 & CCD readout noise & $<15e^{-}$ & $<9e^{-}$  \\
TOM1 & FP mean temperature & $-120<\bar{T}<-80$ C & $-100.15\pm0.05$ C $^{(c)}$\\
TOM2 & FP temperature stability with time & $<\pm0.25$K over 12 h & $<\pm0.15$K over 3 days \\
TOM3 & FP temperature stability across FP & $<10$C  & $<5$C  \\
TOM4 & Dewar vacuum before cooling & $<0.2$mtorr & $<0.01$mtorr\\
TOM5 & Dewar vacuum after cooling & $<0.01$mtorr & $<0.001$mtorr\\
TOM6 & Operation time w/o human intervention & $> 30$h& 1 week\\
TOM8 & FP cooling time & $< 8$h & 6h\\
TOM9 & Dewar warm up time & $< 12 $h & 12h\\
%TOM10 &  FP xy distortion due to vacuum/cooling  & \\
%TOM11 &  FP z distortion due to vacuum/cooling  & \\
 TOM12 &  Vibration due to vacuum/cooling & $<5\mu$m & $<2.5\mu$m\\
 TOM13 & Alignment between FP and C5 & $<30\mu$m & $19\mu$m \\
% TOM19 &  Tolerance of corrector til & \\
% TOM20 &  Stability of tilt angl &\\
 TOM21 &  Horizontal translation accuracy & $<25\mu$m & $<15\mu$m  \\
% TOM22 &  Focus stability over the tilt/translation range\\
% TOM23 & Rotation stability of focus range\\
 TOM24 &  Filter placement accuracy & $<0.5$mm & $<15\mu$m\\
TGFA1 & Guider update rate & $>1$Hz & 1.21Hz \\
TGFA5 & Guide CCD readout noise & $<20e^{-}$ & 9$e^{-}$\\
TGFA6 & Guider region-of-interest  capability & Yes & Yes \\
 \\  %\hline \\
%&\multicolumn{4}{l}{\footnotesize{}}\\
\multicolumn{4}{l}{\footnotesize{(a) Tested at 5 rotation angles (0,45,90,117,131 deg) 
with respect to the optical axis and 2 elevation angles (Zenith and 7 deg).}}\\
\multicolumn{4}{l}{\footnotesize{(b) Tested at environment temperatures between
13 and 38 C and relative air humidity between 15 and 45 percent.}}\\
\multicolumn{4}{l}{\footnotesize{(c) Measured with science grade CCDs after the imager was taken down
from the telescope simulator.}}\\
\\\hline
\end{tabular}
\caption{Technical requirements verified during the telescope simulator tests.}
\label{RequirementsTable}
\end{table}
The technical requirements are divided into groups: General (TG), Optical (TO), CCD detectors (TD), Opto-mechanical (TOM), Guider, Focus and Alignment (TGFA).
The results show that DECam meets all the requirements tested. Other requirements, such as those 
involving optics elements, will be tested at CTIO. Based on these results, the imager was accepted and 
the shipping process has started. All the DECam components are expected to have arrived at CTIO by December 
2011. 

\subsection{Mock observing run}\label{MObs}
By the end of the DECam test cycle on the telescope simulator it became clear that we had
made significant progress towards a fully integrated system. The mock observing run \cite{Soares-Santos:2011a} 
 was a
comprehensive integration test, scheduled for Feb 14-18 2011, in order to verify that
the full system is operational by successfully simulating 4 nights of DES main survey and supernova  
observations. This test allowed us to bring together
new functionalities of SISPI, gather feedback on GUIs, verify that all images produced
have headers useful for analysis, train people to operate the data acquisition system and
identify improvements to be made before we move to CTIO. Observers followed the Observers' Guide,
a document drafted for the mock observing run and will serve as a baseline for the DES 
Observers' Guide.
Of the various components that provide input for SISPI, the Focal Plane was the one we could fully test.
The Data Transfer System was simulated by transferring images from one machine to another in our 
own network, but transfers to remote locations were not tested. 
Other components such as the Telescope Control System and Cloud Camera were not simulated. 
Data Management processing of the images was not a goal for this test.

We operated the system as if we were taking DES data at the mountain top, starting with 
calibration images such as biases,
darks and flats in the various filters and proceeding to science images. We exercised  both
survey and supernova modes using ObsTac, the autopilot responsible for populating the exposure queue according to the survey strategy. We verified that the exposures were taken in survey mode 
when the conditions were set to photometric and that ObsTac would automatically 
switch to supernova mode when conditions
changed to non-photometric, as expected. Image Health, the diagnostics tool, provided basic information about the 
noise level in real time. The images taken generated entries in the Exposure Database Table,
the Electronic Logbook and, in case of problems, the Alarms Database. 
The Guider Control System was used to perform regions-of-interest readouts of the guider chips, but 
we did not test any guiding algorithm because our setup did not simulate the telescope guiding capability.
In addition to the Image Health realtime diagnostics, we tested the Quick Reduce system. Quick Reduce 
processes selected images from the Exposure Table, providing a higher level of diagnostics to the observers.
We exercised the operations model in which we have 2 observers per night. Although experienced observers 
could operate the system without a partner, the 2 observers model has proven to be a 
very efficient way to run the system, allowing for frequent checks of the various diagnostic tools and 
decision making  without unnecessary delay between exposures. 

Overall the system worked very well with only minor glitches from which we could recover 
easily and, although some functionalities were not available for 
testing, we were able to simulate the observing routine and follow the observing plan very closely.
Besides the data that was collected by the SISPI software, the observers were asked to 
provide feedback regarding the use of the software and other aspects of the observing 
in order to provide guidance towards future development. These were gathered into the
mock observing end of run report \cite{Soares-Santos:2011}.

\section{Conclusion}\label{Conclusion}
The installation of DECam at CTIO is imminent. To verify that the DECam imager meets the key  
requirements prior to shipping and in order to minimize the time needed for commissioning we 
built a telescope simulator and performed extensive tests which included a mock observing 
run according to the DES observing strategy. Table \ref{RequirementsTable}
shows that DECam meets all of the requirements tested.  
As we performed a comprehensive test of the system with all the core components 
working together for the first time, we were able to verify the readiness of each 
component and test them in conditions close to what we will experience at the mountain top. 
This allows us to assess the current status and plan ahead for the next stage of the project.  

In the process of performing the mock observing run, 
a  high level of interaction between the SISPI team, the 
DECam team and the collaboration at large was achieved.  
Some of the observers were not familiar with 
the system prior to the mock observing run and were glad to have the opportunity to 
learn through this experience. On the other hand, the SISPI and DECam team benefited 
from this interaction through very positive feedback aimed at the observers needs.   
Also, in the process of preparation for the mock observing run, the overall integration and reliability 
of the system improved at a fast pace. As a result, observers were able to operate the 
system with only minor glitches and by the end of the run, the team was confident that 
commissioning will occur smoothly and DES will be ready to start its first observing season in 
September 2012. 

\section*{Acknowledgements}
Funding for the DES Projects has been provided by the U.S. Department of Energy, the U.S. National Science Foundation, the Ministry of Science and Education of Spain, the Science and Technology Facilities Council of the United Kingdom, the Higher Education Funding Council for England, the National Center for Supercomputing Applications at the University of Illinois at Urbana-Champaign, the Kavli Institute of Cosmological Physics at the University of Chicago, Financiadora de Estudos e Projetos, Funda{\c{c}}\~{a}o Carlos Chagas Filho de Amparo \`a Pesquisa do Estado do Rio de Janeiro, Conselho Nacional de Desenvolvimento Cient{\'{\i}}fico e Tecnol{\'{o}}gico and the Minist\'erio da Ci\^encia e Tecnologia, the Deutsche Forschungsgemeinschaft and the Collaborating Institutions in the Dark Energy Survey.

The Collaborating Institutions are Argonne National Laboratories, the University of California at Santa Cruz, the University of Cambridge, Centro de Investigaciones Energeticas, Medioambientales y Tecnologicas-Madrid, the University of Chicago, University College London, DES-Brazil, Fermilab, the University of Edinburgh, the University of Illinois at Urbana-Champaign, the Institut de Ciencies de l'Espai (IEEC/CSIC), the Institut de Fisica d'Altes Energies, the Lawrence Berkeley National Laboratory, the Ludwig-Maximilians Universit\"at and the associated Excellence Cluster Universe, the University of Michigan, the National Optical Astronomy Observatory, the University of Nottingham, the Ohio State University, the University of Pennsylvania, the University of Portsmouth, SLAC, Stanford University, the University of Sussex, and Texas A \& M University.

%% The Appendices part is started with the command \appendix;
%% appendix sections are then done as normal sections
%% \appendix

%% \section{}
%% \label{}

%% References
%%
%% Following citation commands can be used in the body text:
%% Usage of \cite is as follows:
%%   \cite{key}         ==>>  [#]
%%   \cite[chap. 2]{key} ==>> [#, chap. 2]
%%

%% References with BibTeX database:

\bibliographystyle{elsarticle-num}
\bibliography{bib}

%% Authors are advised to use a BibTeX database file for their reference list.
%% The provided style file elsarticle-num.bst formats references in the required Procedia style

%% For references without a BibTeX database:

% \begin{thebibliography}{00}

%% \bibitem must have the following form:
%%   \bibitem{key}...
%%

% \bibitem{}

% \end{thebibliography}

\end{document}